\documentclass[
aps,
prl,
reprint,
superscriptaddress,
amsmath,amssymb,
notitlepage,
]{revtex4-1}

\usepackage{graphicx}
\usepackage{float}
\usepackage[version=3]{mhchem} 
\usepackage{graphicx}
\usepackage{dcolumn}
\usepackage{bm}
\usepackage{natbib,hyperref}
\hypersetup{colorlinks=true,allcolors=blue}
\usepackage{pdfpages}

\begin{document}

\title{Multiple-pulse lasing from an optically induced harmonic confinement in a highly photoexcited microcavity}

\author{Wei Xie}
\thanks{These authors contributed equally to this work.}
\author{Feng-Kuo Hsu}
\thanks{These authors contributed equally to this work.}
\affiliation{Department of Physics and Astronomy, Michigan State University, East Lansing, MI 48824, USA}
\author{Yi-Shan Lee}
\author{Sheng-Di Lin}
\affiliation{Department of Electronics Engineering, National Chiao Tung University, Hsinchu, Taiwan}
\author{Chih Wei Lai}
\email[]{cwlai@msu.edu}
\affiliation{Department of Physics and Astronomy, Michigan State University, East Lansing, MI 48824, USA}


\begin{abstract}
We report the observation of macroscopic harmonic states in an optically induced confinement in a highly photoexcited semiconductor microcavity at room temperature. The spatially photomodulated refractive index changes result in the visualization of harmonic states in a micrometer-scale optical potential at quantized energies up to 4 meV even in the weak-coupling plasma limit. We characterize the time evolution of the harmonic states directly from the consequent pulse radiation and identify sequential multiple $\sim$10 ps pulse lasing with different emitting angles and frequencies.
\end{abstract}

\maketitle


\begin{figure*}[hbtp]
\centering
\includegraphics[width=0.9\textwidth]{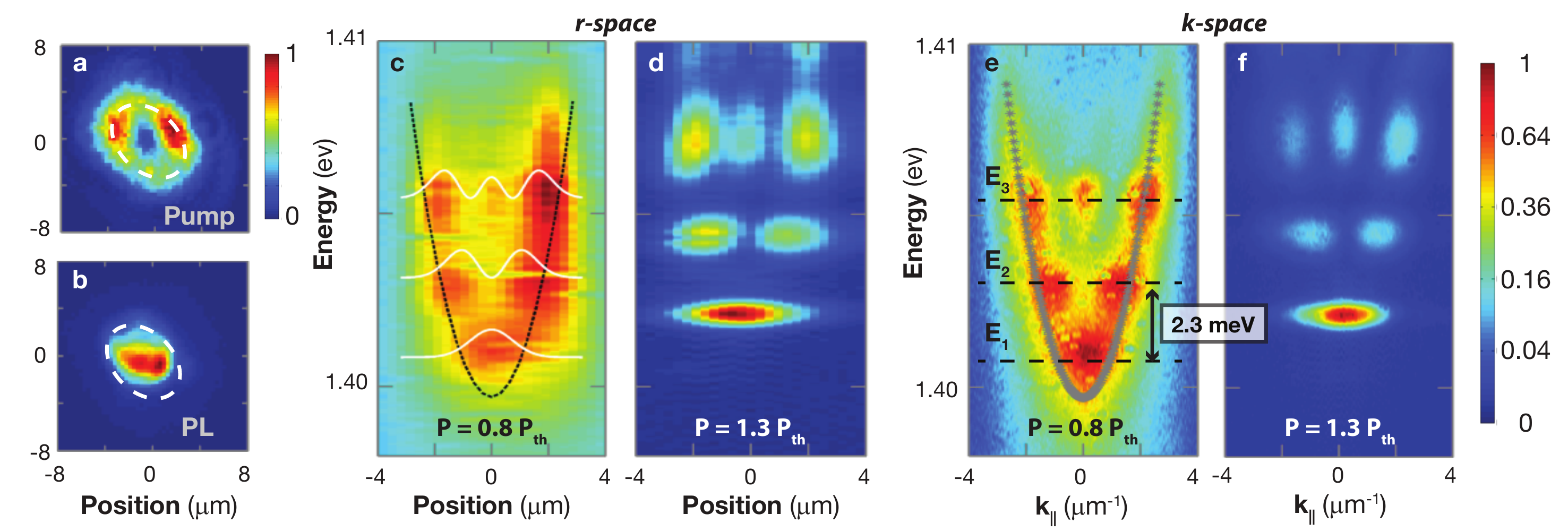}
\caption{\label{fig:rk_spectra}\textbf{Visualization of the macroscopic harmonic states.} (a) Intensity image of the ring-shaped pump laser beam. (b) Photoluminescence (PL) image under a pump flux of about 1.3 $P_{th}$, where the threshold pump flux $P_{th}$ = $1.8\times10^{8}$ photons per pulse. The white dashed line represents the intensity peak of the pump. PL emerges at the center with a minimal overlap with the annular pump laser beam. (c--d) r-space imaging spectra at $P$ = 0.8 $P_{th}$ and 1.3 $P_{th}$. The black dashed line represents the harmonic confining potential $V(x)$, whereas the white lines represent the spatial probability distributions of the lowest three states of a corresponding harmonic oscillator. (e--f) k-space imaging spectra. The energy splitting is $\hbar \omega \approx$ 2 meV, consistent with the quantized energy of a quantum oscillator for a particle with mass $m^* = 3 \times 10^{-5} \ m_e$, as determined by the $E$ vs. $k_\parallel$ dispersion (doted grey line). The quantized modes spectrally blue shift about 1 meV from $P$ = 0.8 to 1.3 $P_{th}$, whereas the quantized energy splitting remains the same. The potential and spectral shifts are due to a density-dependent increase in the chemical potential of the high-density e-h plasma in the reservoir.
}
\end{figure*}

Optical trapping of atoms is essential to realize atomic Bose-Einstein condensates \cite{bagnato1991}. Similarly, the 2D spatial confinement of excitons or exciton-polaritons by inhomogeneous strains \cite{balili2007}, natural defects and potential fluctuations \cite{lai2004,sanvitto2009,degiorgi2014,nguyen2014}, nanofabrication \cite{kaitouni2006,lai2007,bajoni2008,wertz2010,wertz2012,maragkou2010,zajac2012}, and optical potentials \cite{manni2011,tosi2012,cristofolini2013,dreismann2014} in a semiconductor heterostructure/microcavity is considered conducive for the formation and control of the condensates of these quasiparticles. These dynamic condensates can form a meta-stable state in a finite-momentum state \cite{lai2007,nelsen2013} and multiple spatial modes \cite{krizhanovskii2009}. The optical visualization of macroscopic interacting quantum states in solid-state systems has been demonstrated in condensates of excitons (bound electron-hole pairs) and kindred quasiparticles. Such light--matter hybrid condensates are typically formed at cryogenic temperatures in a photoexcited density much below the Mott transition where constituent quasiparticles can be considered boson-like. Optically defined potentials enabled by the effective repulsive interactions of polaritons result in further real-time manipulation of these light-matter fluids in a steady state \cite{tosi2012,cristofolini2013,dreismann2014}. On the other hand, relaxation oscillations with $\sim$10--50 ps period caused by the interplay between reservoir feeding and Bose stimulation have been reported in localized condensates in natural potential fluctuations \cite{degiorgi2014}. 

In this study, we report the optical visualization of dynamic quantized states in a highly photoexcited microcavity at \emph{room temperature}. Photoexcitation creates mainly free pairs of electrons and holes in a III-V-based semiconductor quantum well (QW) as a result of thermal ionization at high temperature \cite{chemla1985,colocci1990a}. In the highly photoexcited microcavity studied here, the nonlinearly photomodulated refractive index results in a sizable effective cavity resonance shift, which enabls optically induced confinement. In an optical confinement initiated by spatially modulated nonresonant ps pulse excitation, sequential multiple-pulse lasing commences at several quantized energy levels. The transverse optical modes in a spatially modulated refractive index can also be understood as an optically induced potential for a fictitious quasiparticle (see Supplemental Material), e.g., correlated \emph{e-h} pairs (\emph{cehp}) resulting from the effective coupling to the cavity light field. Multiple-pulse lasing from discrete states in a harmonic confinement (potential) manifests as a result of the time- and energy-dependent competition between the gain and reservoir carrier relaxation.

The Fabry--P\'erot microcavity sample consists of a $\lambda$ GaAs cavity layer containing three sets of three InGaAs/GaAs QWs embedded within GaAs/AlAs distributed Bragg reflectors (DBRs) (see also Supplemental Material). The sample is nonresonantly excited by a 2-ps pulse pump laser at $E_p$ = 1.58 eV at room temperature. The pump energy is about 250 meV above the QW band gap ($E_g' \approx$ 1.33 eV) and 170 meV above the cavity resonance ($E_c \approx $ 1.40--1.41 eV). The high-density \emph{e-h} plasmas of a density $\approx1-5 \times 10^{12} \, \text{cm}^{-2}$ per QW per pulse are formed momentarily after \emph{non-resonant} pulse excitation as a result of rapid ($<$10 ps) energy dissipation through optical phonons. The radiative recombination rate of these \emph{e-h} carriers in the reservoir is suppressed because the cavity resonance $E_c$ is detuned to $\sim$ 70--80 meV above the QW bandgap $E_g'$, i.e., the $e1hh1$ transition between the first quantized electron and (heavy-)hole states in a QW. Below the lasing threshold, the high-density \emph{e-h} plasmas in the microcavity are subject to non-radiative loss with a long decay time ($>$500 ps) (Supplemental Fig.~S1). Therefore, the chemical potential of the \emph{e-h} plasmas ($\mu_0$) appears to be stationary within $\sim$100 ps after pulse excitation. The bare cavity resonance $E_c$ of the sample studied here is close to $E_g''$, the $e2hh2$ transition between the second quantized electron and hole levels of InGaAs/GaAs QWs.

When $\mu_0$ advances toward $E_c$, the average refractive index near $E_c$ can be considerably modified, and this results in a sizable blueshift of the effective cavity resonance ($E'_c$) and consequent emission energy (Supplemental Fig.~S2a). In general, $E'_c$ increases with the photoexcited density and can be seen as an effective potential $V$ for \emph{cehp}s (Supplemental Material section S3). Therefore, a quasi-stationary confining potential $V(r)$ for \emph{cehps} can be established by a ring-shaped spatial distribution of photoexcited carrier density. In our experiments, we use a \emph{double-hump-shaped} beam to fixate the orientation of the optically defined potential (Fig. \ref{fig:rk_spectra}a--b).

Real-space (r-space) imaging spectra provide direct visualization of the optical potential. Fig.~\ref{fig:rk_spectra}c--d shows the r-space imaging spectra of a narrow cross-sectional stripe across the trap. A nearly parabolic potential well of $\sim$10 meV across 3 $\mu$m is revealed. Such a quasi-1D harmonic potential, $V(x) =  1/2 \ \alpha x^2$, is spontaneously formed under 2 ps pulse excitation. The resultant radiation appears in-between the two humps, a few micrometers away from the pump spot. The r-space intensity profiles agree with the probability distributions of the quantized states of a harmonic oscillator (harmonic states). The deviations of the actual potential from a perfectly harmonic trap result in slightly asymmetric luminescence intensity distributions. These standing wave patterns form in the self-induced harmonic optical potential when a macroscopic coherent state emerges. At a critical density, the occupation number of \emph{cehp}s ($n_i$) in these quantized harmonic states approaches unity when the conversion from the reservoir ($N_{eh}$) overcomes the decay of the harmonic states. Consequently, when a stimulated process ($\propto N_{eh} n_i$) prevails, $n_i$ and the resultant radiation increase nonlinearly by a few orders of magnitude with the increasing $N_{eh}$.

\begin{figure}[htb]
\centering
\includegraphics[width= 0.5 \textwidth]{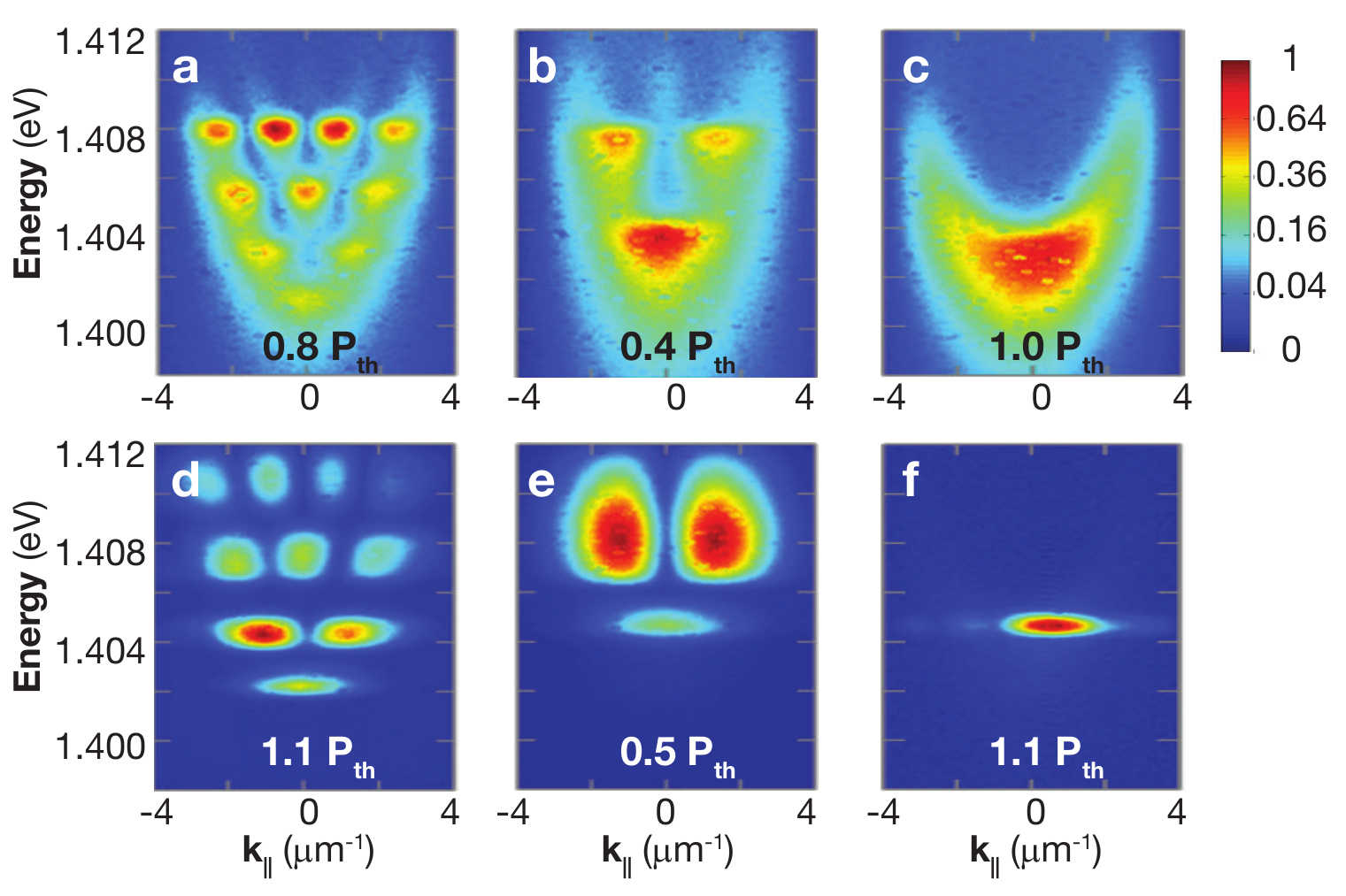}
\caption{\label{fig:ek_size}\textbf{Quantized states in optically controlled confining potentials.} K-space imaging spectra under below-threshold (a--b) and above-threshold (d--e) for two double-hump-shaped pump beams with peak-to-peak distances of 5 $\mu$m and 3 $\mu$m, respectively. For comparison, the k-space imaging spectra under a flat-top pump beam are shown in (c) and (f). The corresponding r-space images and spectra are shown in Supplemental Fig. S1.
}
\end{figure}

\begin{figure*}[htb]
\centering
\includegraphics[width=0.6\textwidth]{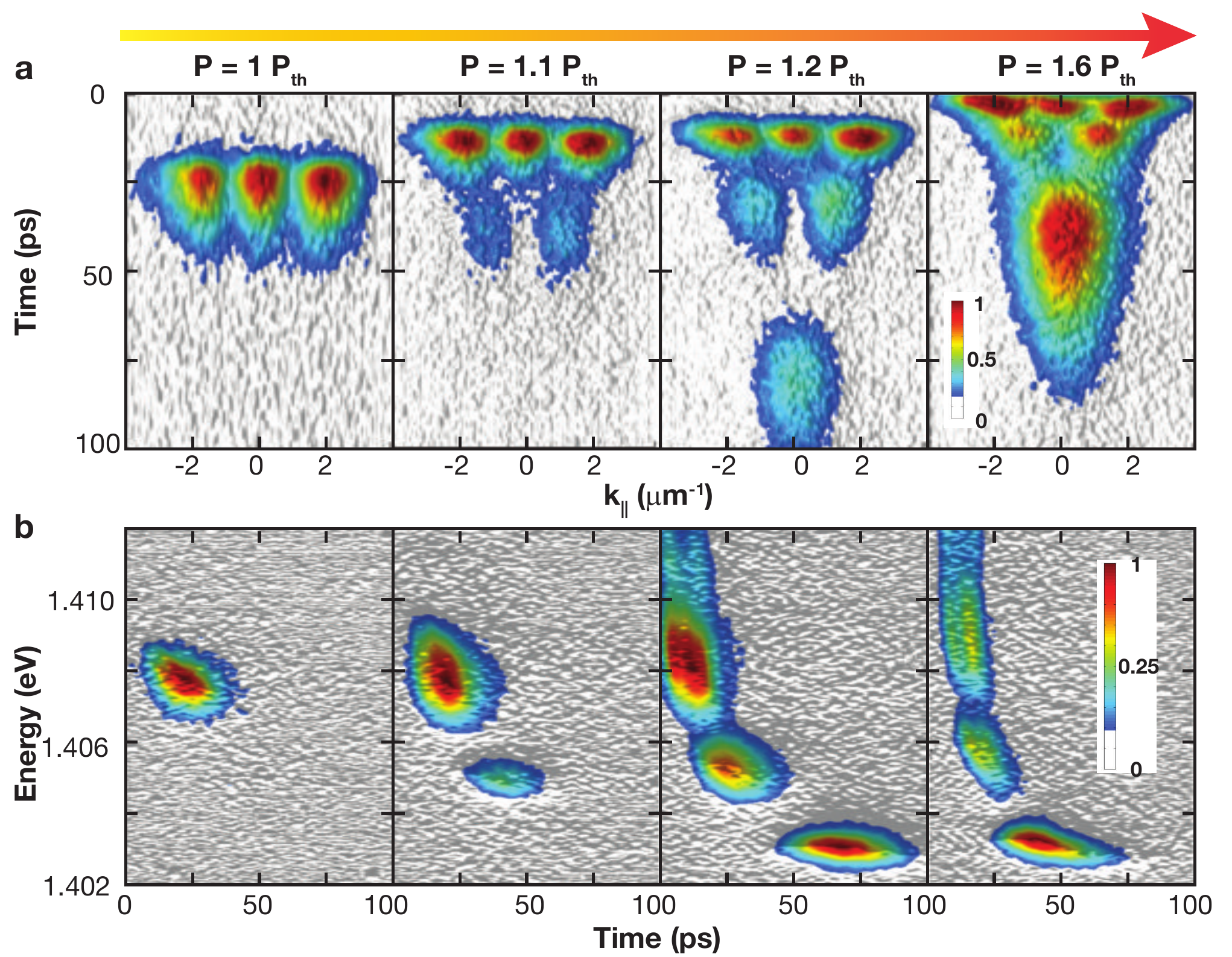}
\caption{\label{fig:dynamics}\textbf{Dynamics.} (a) Time-dependent luminescence in k-space at $P = 1.0 $, $1.1$, $1.2$ and $1.6 \, P_{th}$. The $E_3$, $E_2$ and $E_1$ states appear sequentially with the increasing pump flux. The rise times decrease with the increasing pump flux for all states. (b) Time-dependent spectra in r-space. The false color represents normalized intensities.
}
\end{figure*}

An even more regular pattern appears in k-space imaging spectra (Fig.~\ref{fig:rk_spectra}e--f). The $E$ vs. $k_\parallel$ dispersion measured below the threshold allows the direct measurement of the effective mass $m^{*} = 3\times 10^{5} \, m_e$, where $m_e$ is the electron rest mass. Moreover, the strength of the optically defined harmonic potential $\alpha$ can be tuned through a variation in the spatial distance between two humps of the pump beam (Fig.~\ref{fig:ek_size}). The energy quantization ($\hbar \omega$) varies with $\sqrt{\alpha/m^*}$, while the emission patterns evolve into the probability distributions for a particle with an effective mass $m^*$ in $V(x)$ (Fig.~\ref{fig:ek_size} and Supplemental Fig.~S3).

\begin{figure}[htb]
\centering
\includegraphics[width=0.35\textwidth]{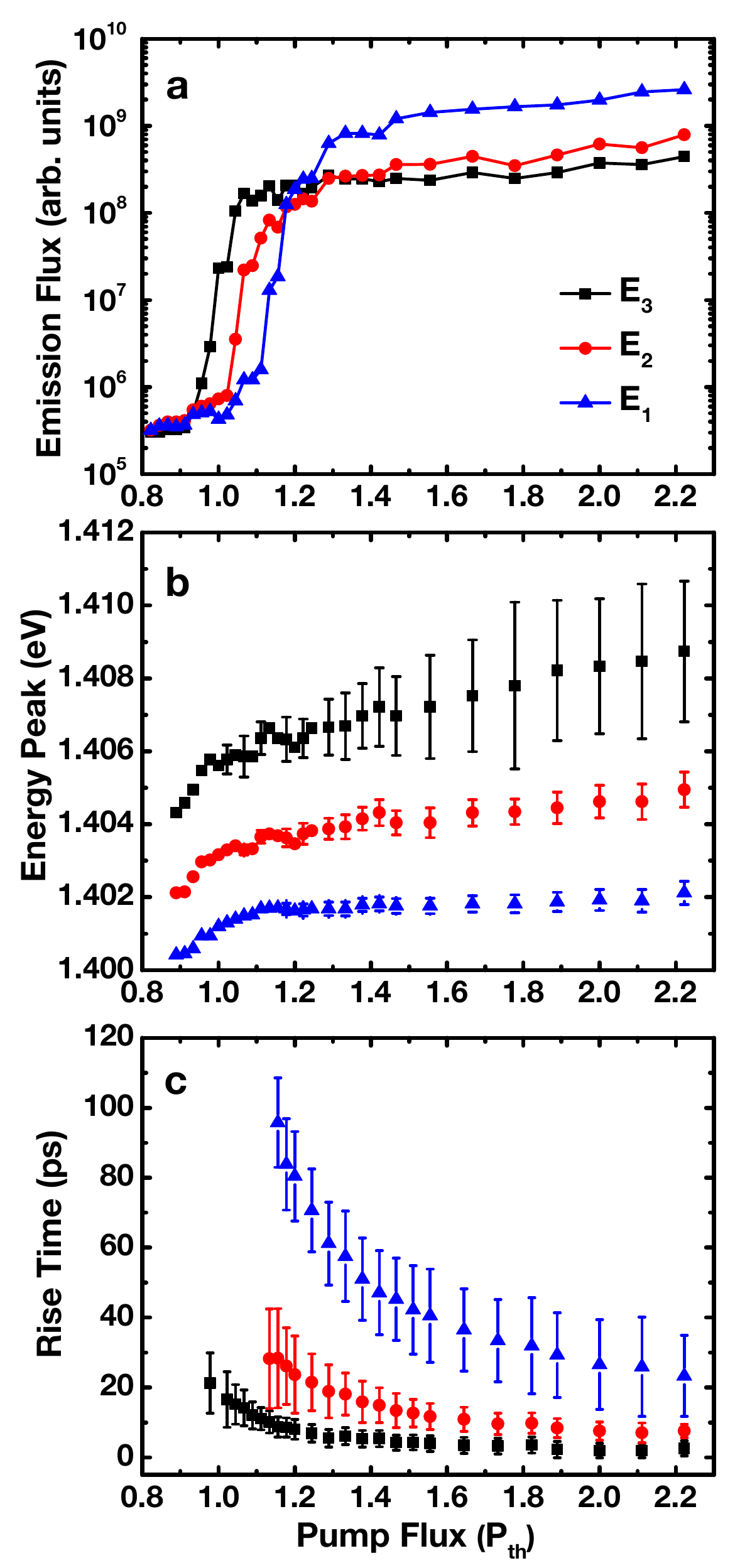}
\caption{\label{fig:density}\textbf{Density dependence.} (a) Temporally and spectrally integrated emission flux vs. pump flux. All three modes display non-linear increases in intensity by more than two orders of magnitude, and saturate at $1.1$, $1.2$ and $1.3 \, P_{th}$, respectively. (b) Peak energy (solid shapes) and linewidths 2$\Delta E$ (error bars) vs. pump flux. These states spectrally blue shift by 1 to 4 meV. The spectral linewidths ($\Delta E$) and pulsewidths ($\Delta t$) are reciprocal with a product of $\Delta E \times \Delta t \approx 4 \hbar$ ($\hbar$) for $E_3$ and $E_1$ ($E_2$), which is closed to the uncertainty (Fourier-transform) limit. (c) Rise time vs. pump flux for the three states $E_1$ (blue), $E_2$ (red) and $E_3$ (black). The error bar represents 2$\Delta t$.}
\end{figure}

Temporally, these harmonic states emerge sequentially and display distinct density-dependent dynamics. In Fig. \ref{fig:dynamics}a, we study the time evolutions of three harmonic states in k-space. The corresponding time-integrated imaging spectra in k-space are shown in Supplemental Fig.~S4. The energy relaxation of these three states is revealed in the time-resolved spectra in Fig.~\ref{fig:dynamics}b. At a critical photoexcited density, the high-energy $E_3$ state arises $\sim$ 25 ps after pulse excitation and lasts for $\sim$ 20 ps. The corresponding pump flux is defined as the threshold ($P_{th}$). With the increasing pump flux, the $E_2$ state emerges $\sim$ 25 ps after $E_3$ at 1.1 $P_{th}$, whereas the \emph{ground} $E_1$ state appear 50 ps after $E_2$ at 1.2 $P_{th}$. In the optically induced harmonic confinement studied here, the effective cavity resonance $E'_c$ decreases with time as a result of the decay of reservoir carriers. However, the conversion from the reservoir to specific confined $E_i$ state can be efficient only when $E'_c$ is resonant with $E_i$. Such a temporally decrease in $E'_c$ results in a time- and energy-dependent conversion efficiency for these confined harmonic states (Fig. S1) and consequent multiple-pulse lasing above the critical density threshold.

We further analyze the emission flux and energy of these harmonic states by using time-integrated spectra measured with increasing photoexcited density (Fig. \ref{fig:density} and Supplemental Fig.~S2). Far below the threshold ($P < 0.4 \ P_{th}$), emission is dominated by luminescence from GaAs spacer layers. When the pump flux is increased, the emissions from the \emph{cehp} states become increasingly dominant, and the $E_3$ state eventually lases at the threshold. The emission fluxes of all three states increase nonlinearly by more than two orders of magnitude across a threshold and then reach a plateau at a saturation density (Fig.~\ref{fig:density}a). On the other hand, the emission energy of these three states increases to a constant with the increasing pump flux (Fig.~\ref{fig:density}b). The energy spacing $\hbar \omega$ only increases slightly with density. The spectral linewidths increase slightly for the $E_1$ state but by about a factor of 10 for the $E_3$ state. Next, we study the density-dependent dynamics. Fig.~\ref{fig:density}c shows the rise times and pulse durations for $E_3$, $E_2$, and $E_1$. The product of the variances of the spectral linewidth ($\Delta E$) and the pulse duration ($\Delta t$) is found to be close to that of a transform-limited pulse: $\gtrsim 4 \hbar$ and $\approx 1 \hbar$ for the $E_3/E_1$ and $E_2$ states, respectively. These harmonic states are macroscopically coherent states with finite phase and intensity fluctuations induced by interactions.

We use a rate-equation model to describe the dynamic formation of quantized states in an optically defined harmonic confinment (Supplemental Material Sec. S4Supplemental Material). This phenomenological model reproduces qualitatively the dynamics and integrated emission flux of the harmonic states when photoexcited density is varied (Supplemental Figs.~S5, S6, and S7). Non-equilibrium polariton condensates have been modeled by a modified Gross--Pitaevskii (GP) (or complex Ginzberg--Landau [cGL]) equation that accounts for the finite lifetime of polaritons \cite{szymanska2006,wouters2007}. However, cGL-type equations are inapplicable for the multiple dynamic states examined in this study. Additionally, the formation of a BEC-like condensate that underpins the GP- or cGL-type equation is not necessarily justified in our room-temperature experiments.

Transverse light-field patterns and confined optical modes have been identified in nonlinear optical systems \cite{mandel2004}, vertical-cavity surface-emitting lasers (VCSELs) \cite{chang-hasnain1991,zhang2004}, and microscale photonic structures \cite{reithmaier1997}. In principle, the multiple transverse mode lasing in a high-density \emph{e-h}-plasma described in the present study can be modeled by a self-consistent numerical analysis with Maxwell-Bloch equations developed for conventional semiconductor lasers \cite{koch1995,sarzala2012}, provided that the strong optical nonlinearities induced by Coulomb many-body effects, such as screening, bandgap renormalization, and phase-space filling, are all considered. For example, one can consider the formation of index-guided multiple transverse modes as a result of an optically-induced refractive index reduction ($\Delta n_c(x)$). The cavity resonance shift ($\delta E$) can be estimated from $\delta E_c/E_c = - \Delta n_c/n_c$, where $n_c$ is the effective refractive index averaged over the longitudinal cavity photon mode which spans over $\sim\mu$m in the growth direction. Therefore, a cavity resonance shift $\delta E \sim$10 meV (Figs. \ref{fig:rk_spectra} and \ref{fig:ek_size}) corresponds to $|\Delta n_c/n_c| \sim$ 1\%. Such a significant refractive index is probable with resonance-enhance optical nonlinearity from the \emph{e-h} correlation \cite{schmitt-rink1986a,campi1998,tanguy1999,kamide2012} or carrier-induced change in refractive index at a high carrier density (of $\sim10^{19}$ cm$^{-3}$ or more) \cite{lee1986,bennett1990,huang1998}. To uncover the microscopic formation mechanisms of such a sizable spatial modulation in refractive index or equivalent effective harmonic confinement under pulse excitation, further characterizing the photoexcited density distribution with the use of other ultrafast spectroscopic techniques, such as a pump-probe spectroscopy, is necessary.

We identify sequential multiple $\sim$10 ps pulse lasing in an optically induced harmonic confinement in a semiconductor microcavity at room temperature. Laser radiation emerges at the quantized states of an optically induced harmonic confinement. The lasing frequency, rise time, pulse width, and radiation angle can be controlled through a variation in the photoexcited density or optical pump spot dimensions in real time. The sample has a composition structure similar to those used for studies of polariton condensates/lasers and the widely used VCSELs. In the highly photoexcited microcavity studied here, harmonic confinement can be optically induced and controlled as a result of photomodulated refractive index changes even in the weak-coupling plasma limit. Our demonstration of macroscopic harmonic states in a room-temperature optically induced harmonic confinement improves our understanding of nonlinear laser dynamics and should stimulate studies on emergent ordered states near the Fermi edge of a high-density \emph{e-h} plasma in a semiconductor cavity \cite{nozieres1985,keeling2005,yamaguchi2013}.

\begin{acknowledgments}
We thank Cheng Chin, Mark Dykman, Brage Golding, Peter B. Littlewood, John A. McGuire, David W. Snoke and Carlo Piermarocchi for the discussions. This work was supported by NSF grant DMR-0955944 and J. Cowen endowment at Michigan State University.	
\end{acknowledgments}

\bibliography{/Users/cwlai/Dropbox/Bib/lai_lib}
\clearpage


\section*{Supplementary material}

# Supplemental Material for
# Multiple-pulse lasing from an optically induced harmonic confinement in a highly photoexcited microcavity

Wei Xie,[1] Feng-Kuo Hsu,[1] Yi-Shan Lee,[2] Sheng-Di Lin,[2] and Chih-Wei Lai[1]

[1]*Department of Physics and Astronomy, Michigan State University, East Lansing, MI 48824, USA*

[2]*Department of Electronics Engineering, National Chiao Tung University, Hsinchu 30010, Taiwan*

## S.1. MATERIALS AND METHODS

**Sample.** The microcavity sample is grown on a semi-insulating (100)-GaAs substrate with the molecular beam epitaxy method. The top (bottom) DBR consists of 17 (20) pairs of GaAs(61-nm)/AlAs (78-nm) $\lambda/4$ layers. The central cavity layer consists of three sets of three $In_{0.15}Ga_{0.85}As$/GaAs (6 nm/12 nm) quantum wells each, positioned at the anti-nodes of the cavity light field. The structure is undoped and contains a $\lambda$ GaAs cavity sandwiched by DBRs. The bare cavity resonance $E_c \approx 1.40$ eV at room temperature. The QW bandgap ($E'_g \sim 1.33$ eV) is tuned through a rapid thermal annealing process (at 1010°C-1090°C for 5-10 s), in which the InGaAs QW bandgap blueshifts because of the diffusion of gallium ions into the MQW layers. With increasing photoexcited density, the chemical potential $\mu_0$ of the $e$-$h$ plasma can increase up to $\sim$80 meV above $E'_g$. The cavity quality factor Q is about 4000–7000, corresponding to a cavity photon lifetime of $\sim$2 ps.

**Optical excitation.** The front surface of the sample is positioned at the focal plane of a high-numerical-aperture objective (N.A. = 0.42, 50×, effective focal length: 4 mm). The holographic beam shaping is achieved with a reflected Fourier transform imaging system that consists of a 2D phase-only spatial light modulator (SLM), a 3× telescope, a Faraday rotator, a polarizing beam splitter, and the objective. The light fields at the SLM and the sample surface form a Fourier transform pair. The 2D SLM (1920 × 1080 pixels, pixel pitch = 8 $\mu$m) enables us to generate arbitrary pump geometries with a $\approx$2 $\mu$m spatial resolution at the sample surface by using computer-generated phase patterns. The polarization properties of pump and luminescence are controlled/analyzed with liquid-crystal devices. The pump flux is varied by more than two orders of magnitude with a liquid-crystal-based attenuator.

**Thermal management.** At a high pump flux, the steady-state incident power transmitted to the sample at the 76 MHz repetition rate of the laser would exceed 50 mW, and significant thermal heating results. Thermal heating can inhibit laser action and lead to spectrally broad redshifted luminescence. To suppress thermal heating, we temporally modulate the 2 ps 76 MHz pump laser pulse train with a duty cycle (on/off ratio) < 0.5% by using a double-pass acousto-optic modulator system. The time-averaged power is limited to below 1 mW for all experiments.

**R-space and k-space imaging spectroscopy.** We measure the angular, spectral, and temporal properties of luminescence in the reflection geometry. The angle-resolved (k-space) luminescence images and spectra are measured through a Fourier transform optical system. A removable $f = 200$ mm lens enables the projection of either the k-space or r-space luminescence on to the entrance plane or slit of the spectrometer. Luminescence is collected through the objective, separated from the reflected specular and scattered pump laser light with a notch filter, and then directed to an imaging spectrometer. A lasing mode with a spatial diameter $\approx$8 $\mu$m is isolated for measurements through a pinhole positioned at the conjugate image plane of the microcavity sample surface. The spectral resolution is $\approx$0.1 nm (150 $\mu$eV), which is determined by the dispersion of the grating (1200 grooves/mm) and the entrance slit width (100–200 $\mu$m). The spatial (angular) resolution is $\approx$ 0.3 $\mu$m (6 mrad) per CCD pixel.

## S.2. SUPPLEMENTARY FIGURES

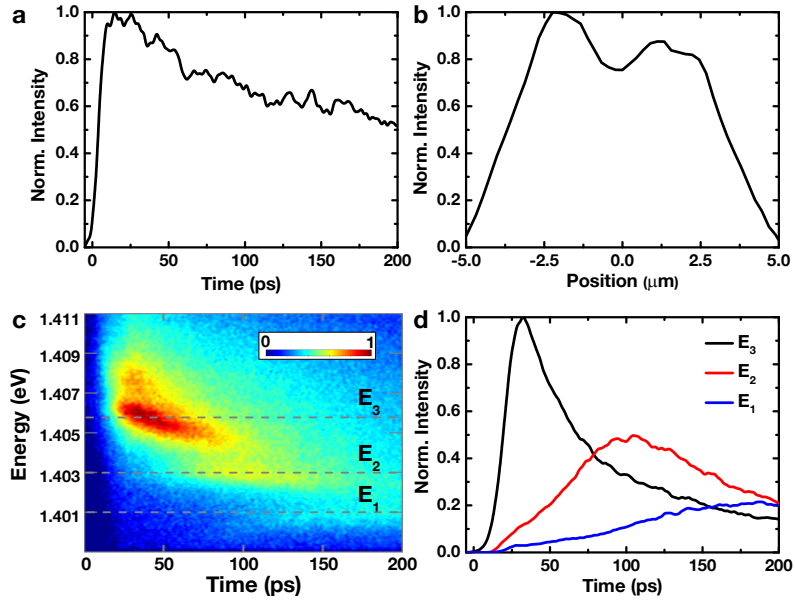

Fig. S1. **Below-threshold dynamics.** (a) Time-dependent spectrally and spatially integrated luminescence at $P = 0.9\ P_{th}$. The luminescence decays with a $> 500$ ps time constant largely because of the nonradiative loss. (b) Spatial luminescence distribution integrated up to a delay of 50 ps after pulse excitation, revealing a double-hump-shaped profile. (c) Time-dependent luminescence spectra at $P = 0.9\ P_{th}$. The false color represents the square root of the luminescence intensity. (d) Time-dependent luminescence intensity for the $E_3$ (black), $E_2$ (red) and $E_1$ (blue) harmonic states. The intensity is integrated over a spectral range of 1 meV centering the energies, as indicated by the dashed lines in (c).

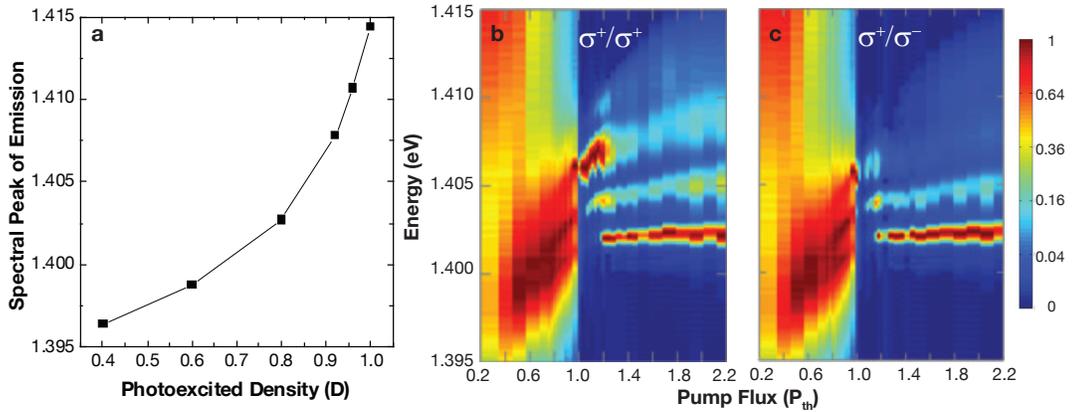

Fig. S2. **Pump-flux-dependent resonance energy and polarized luminescence spectra.** (a) Resonance energy (the energy of emission mode at $k_\parallel \approx 0$ in K-space spectral mapping) versus photoexcited density under a tightly-focused circular pumping beam (radius $\sim 1\ \mu$m). The spectral peak can be considered as the effective cavity resonance and displays a sizable spectral blueshift with increasing the photoexcited density. Such an energy shift with photoexcited density is the basis of optically induced harmonic confinement (potential) using a spatially modulated pump beam profile. $D \approx 4 \times 10^{12}\ \text{cm}^{-2}$ per quantum well per pulse. (b,c) Normalized spectra integrated over k-space as a function of pump flux in the double-hump-shaped pumping case: Co-circularly (b) and cross-circularly (c) polarized component. Quantized harmonic states are formed above $\sim 0.8 P_{th}$. Below the threshold, the luminescence from all states is unpolarized. Above the threshold, the $E_3$ state is highly circularly polarized, where as the $E_1$ state has a diminishing circular polarization. The fourth quantized state ($E_4$) is weakly confined.

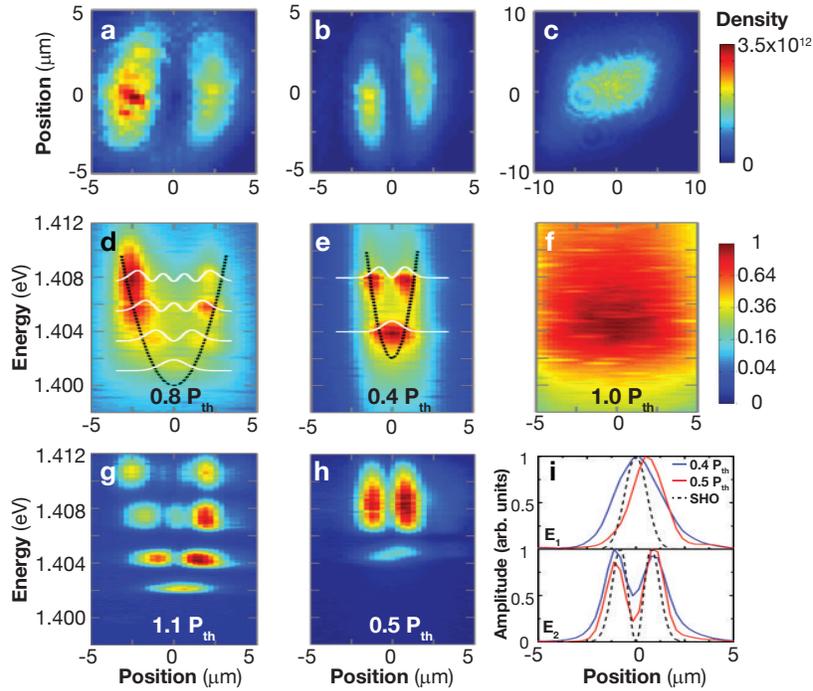

Fig. S3. **Quantized states vs. the trap size.** (a–c) Optical images of pump beam profiles at the front sample surface for two double-hump-shaped and one flat-top focal spots. The false color represents the photoexcited density ($cm^{-2}$) per quantum well per pulse at 0.8 $P_{th}$ for (a) and at 0.4 $P_{th}$ for (b). $P_{th} = 1.8 \times 10^8$ per pulse, as defined in the main text. The lasing thresholds are 0.8 $P_{th}$ for (a) and 0.4 $P_{th}$ for (b). (d–f) Real-space imaging spectra measured below the threshold. The black dashed line shows the effective harmonic potential, whereas the white lines represent the spatial probability distributions of the lowest few states of a corresponding quantum oscillator. The quantized energy splittings for (d) and (e) are 2.2 and 4.0 meV, respectively. The false color represents the square root of intensity. (g–h) Real-space imaging spectra under above-threshold pump. (i) The measured wave functions of the two lowest states $E_2$ and $E_1$ at 0.4 $P_{th}$ (solid blue lines) and 0.5 $P_{th}$ (solid red lines) as shown in (e) and (h). The ideal single-particle wave functions of the corresponding quantum simple harmonic oscillator (SHO) are represented by the dashed lines.

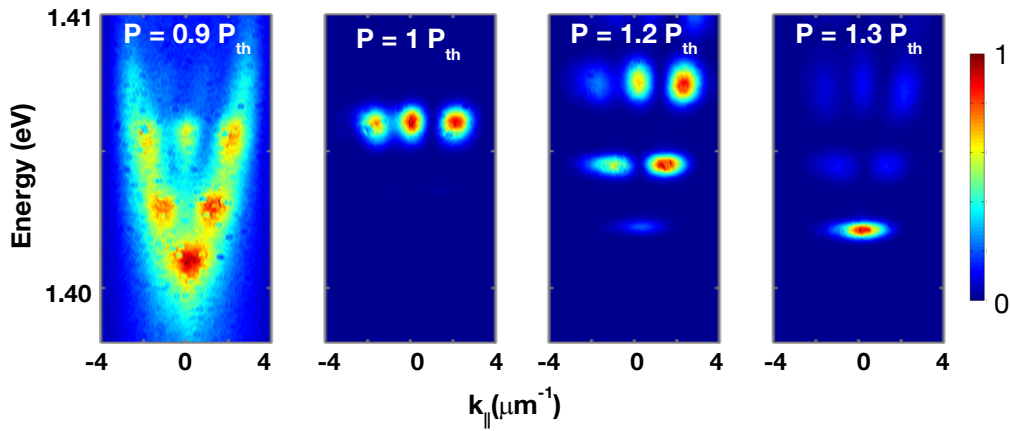

Fig. S4. **K-space imaging spectra.** The energy vs. in-plane momentum $k_{||}$ dispersion corresponding to the data set in the main paper with the increasing pump flux from 0.9 to 1.3 $P_{th}$. The quantized $E_3$, $E_2$, and $E_1$ states lase in sequence with the increasing pump flux. $E_3$-state lasing commences first, followed by $E_2$-state lasing at about 1.2 $P_{th}$, and then $E_1$-state lasing appears at 1.3 $P_{th}$ and dominates over the other two states at a higher pump flux.

### S.3. OPTICALLY INDUCED POTENTIAL AND REFRACTIVE INDEX CHANGES

The transverse optical modes confined by a spatially modulated refractive index can also be understood as an optically induced potential for a fictitious quasiparticle.

Considering the electromagnetic fields in a planar Fabry-Pérot cavity with a medium of a 2D spatially modulated effective refractive index profile $n(r)$, we can express the Helmholtz equation for the electric field in the following simplified form:

$$\frac{\partial^2 F(r,\varphi,z)}{\partial r^2} + n^2(r) \cdot k_0^2 \cdot F(r,\varphi,z) = 0 \tag{1}$$

The electric field F in the cavity can be decomposed into lateral and transverse components, and approximated as $F(r,\varphi,z) \propto \psi(r)\cdot \exp[-i\, n(r)\, k_z(r)\cdot z)]$, where $k_0^2 = k_z^2(r) + k_\parallel^2(r)$, and $k_\parallel(r)$ and $k_z(r)$ are transverse and longitudinal wavenumbers in the vacuum, respectively. For simplicity, we consider the 1D case and reduce Eq. 1 to:

$$\frac{\partial^2 F(x,z)}{\partial x^2} + n^2(x) \cdot [k_0^2 - k_z^2(x)] \cdot F(x,z) = 0 \tag{2}$$

Applying the effective-mass approximation for small $k_\parallel$, we express the in-plane dispersion $E_c(k_\parallel)$ in terms of an effective mass $m^*$ for the excitation: $E_c(k_\parallel) = E_{c0} + E_\parallel = \frac{(n\hbar k_z)^2}{2m^*} + \frac{(n\hbar k_\parallel)^2}{2m^*}$, where $m^* \equiv \frac{\hbar^2 n^2 k_z^2}{2E_{c0}} = \frac{\hbar n^2 k_z}{2c}$.

Eq. 2 can then be mapped to a time-independent Schrödinger equation by replacing the electric field amplitude with the probability wavefunction $\phi$ of a hypothetical quasi-particle with an effective mass $m^*$ in the presence of a potential $V(x)$, thus:

$$\frac{-\hbar^2}{2m^*}\cdot \frac{\partial^2 \phi(x)}{\partial x^2} + [V(x) - E_c]\cdot \phi(x) = 0 \tag{3}$$

Replacing the effective mass for an optical cavity mode with longitudinal wavenumber $k_z$, we obtain

$$\frac{\partial^2 \phi(x)}{\partial x^2} + \frac{n^2 k_z}{\hbar c}\cdot [E_c - V(x)]\cdot \phi(x) = 0 \tag{4}$$

Comparing Eqs. 2 and 4 gives

$$V(x) = \hbar c \cdot k_z(x) = E_{c0}(x) \tag{5}$$

Therefore, $E_{c0}(x)$ and $E_\parallel$ correspond to the potential $V(x)$ and the kinetic energy of a quasi-particle of mass $m^*$. Considering $E_{c0}(x) \propto \frac{\hbar c}{n(x) L_c}$, where $L_c$ is the effective cavity length, we can relate the potential to the spatially modulated refractive index $\frac{\Delta n(x)}{\bar{n}}$ as follows: $V(x) = \bar{V}\cdot [1 + \frac{\Delta V(x)}{\bar{V}}] = \frac{\overline{E}_{c0}}{1+\Delta n(x)/\bar{n}} \approx \overline{E}_{c0}\left(1 - \frac{\Delta n(x)}{\bar{n}}\right)$, where $\bar{n}$ and $\bar{V}$ are the refractive index and potential at $x = 0$. Thus, the spatial modulation of the potential ($\Delta V(x)$) and refractive index ($\Delta n(x)$) have a one-to-one correspondence: $\frac{\Delta V(x)}{\bar{V}} \approx -\frac{\Delta n(x)}{\bar{n}}$.

### S.4. THEORETICAL MODELING

As discussed in section S.3, the spatial modulation of the refractive index can be considered a spatially dependent effective potential for a quasiparticle with an effective mass associated with the bare cavity resonance and mean refractive index. The multiple transverse optical modes are equivalent to harmonic states in harmonic confinement (potential). In this section, we use a set of rate equations that consider the formation of macroscopic harmonic states via the stimulation of a fictitious quasiparticle (correlated e-h pairs [cehps]) in the presence of an effective harmonic potential. We consider the temporal evolution of the reservoir distribution $N_R(x,t)$ and quantized sates $n_i(t)$. The e-h pairs photoexcited nonresonantly by a 2 ps pulse laser cool down rapidly (<5 ps) to the band edge. The cooled carriers in the reservoir $N_R(x,t)$ are subject to nonradiative loss ($\Gamma_{nr}$) and thus result in a slow decrease (~0.1 meV/ps) in the chemical potential $\mu_0$ of the reservoir carriers. In this model, the bare cavity resonance $E_c$ is fixed. By contrast, the effective cavity resonance $E_c'(t)$ blueshifts with the increasing photoexcited density (Fig. S2) and redshifts with time as a result of the temporal decay of reservoir carriers (Fig. S5). Standing waves (macroscopic coherent states of cehps) are formed in an optically induced harmonic confinement (potential $V(x) \propto x^2$) associated with photomodulated refractive index changes initiated by a double-humped pump beam profile. Such a harmonic potential is quasi-stationary within ~100 ps after the pulse excitation as a result of ~ns decay of the reservoir carriers.

A fraction of the reservoir carriers $[N_{eh}(x,t) = \beta N_R(x,t)]$ near the effective cavity resonance $E'_c$ are considered as *cehp*s that can be scattered into the harmonic states via a spontaneous or stimulated process. Those *cehp*s with energy nearly in resonance with a specific $E_i$ state can be spontaneously scattered into the $E_i$ state at a rate $W_s$. Above the threshold, the population of $E_i$ states increases nonlinearly as a result of stimulation ($\propto W_{ss} N_{eh} n_i$). The consequent leakage photons from the decay of these confined *cehp*s at a rate associated with the cavity photon decay rate $\Gamma_c$ are then measured experimentally. The dynamics of the quantum oscillator of *cehp*s can then be described by the following set of coupled rate equations:

$$\frac{d N_R(x,t)}{dt} = G_p(x,t)P - \Gamma_{nr}\, N_R(x,t) - \sum_{i=1}^{3} \eta_i\, \zeta_i\, N_{eh}(x,t)\, [W_{ss}\, n_i(t) + W_s], \tag{6}$$

$$\frac{d n_i(t)}{dt} = \int dx\, \eta_i\, \zeta_i\, N_{eh}(x,t)\, [W_{ss}\, n_i(t) + W_s] - \Gamma_c\, n_i(t). \tag{7}$$

The spatial generation rate $G(x,t)$ follows the 2 ps Gaussian temporal and double-hump-shaped beam profile (Fig. S5c and Fig. S6c). $P$ is the pump flux. $\zeta_i$ is a scaling factor associated with $\beta$ for a specific $E_i$ state. $\eta_i = \exp\left[-(E'_c - E_i)^2/(2\,\Delta E_i^2)\right]$ is the coupling efficiency of the reservoir *cehp*s to the $E_i$ state, where $\Delta E_i$ is the spectral linewidth of the $E_i$ state.

The fitting parameters are as follows: $\beta = 0.0125$, $\Gamma_{nr} = 1/700$ ps$^{-1}$, $\Gamma_c = 0.5$ ps$^{-1}$, $W_s = 1.0 \times 10^{-4}$ ps$^{-1}$, $W_{ss} = 1.0 \times 10^{-2}$ ps$^{-1}$, $\zeta_1 = 0.55$, $\zeta_2 = 0.67$, $\zeta_3 = 0.80$, $\Delta E_i = \Delta E = 1$ meV, $E_1 = 1.401$ eV, $E_2 = 1.403$ eV, and $E_3 = 1.405$ eV. In this simulation, we neglect the spin degree of freedom and only consider the density regime above 0.8 $N_R^{th}$ and $|E'_c(x,t) - E_i| < 15$ meV ($N_R^{th} \sim 500$, corresponding to the occupation number at $P = P_{th}$). We approximate $E'_c(x,t) = E_{c0} + g\, N_{eh}(x,t)$, where $E_{c0}$ and $g$ are determined by the following initial conditions: $E'_c(x=0, t=0) = 1.4$ eV and $E'_c(x = \pm 3\mu m, t=0) = 1.407$ eV.

This model reproduces the time evolution of the quantized states with increasing photoexcited density (Fig. S6). Below the threshold, the scattering of the reservoir *cehp*s into the harmonic states occurs through a spontaneous decay process. The radiative quantum efficiency is low ($< 10^{-4}$) because of the dominant non-radiative recombination loss. Above the threshold, the quantum efficiency increases by two to three orders of magnitude as a result of stimulation. This model reproduces qualitatively the rise times and the time-integrated emission fluxes of the multiple pulse lasing from the quantized states when the pump flux is varied (Fig. S7).

We note a few shortcomings of this simplified model: (1) The model does not consider the local energy shift and the evolution of potentials induced by the interactions of the *cehp*s because the density-dependent $E'_c(N_R, N_{eh}, n_i)$ is not well determined. As a result, the density-dependent energy shifts and linewidths cannot be reproduced. (2) The model does not consider carrier diffusion, which affects dynamics at a density far above the threshold. (3) Phase and intensity fluctuations are neglected.


\end{document}